\documentstyle[aps]{revtex}
\begin{document}
\newcommand{\beq}{\begin{equation}}
\newcommand{\eeq}{\end{equation}}
\newcommand{\beqn}{\begin{eqnarray}}
\newcommand{\eeqn}{\end{eqnarray}}
\newcommand{\bmath}{\begin{mathletters}}
\newcommand{\emath}{\end{mathletters}}
\twocolumn[\hsize\textwidth\columnwidth\hsize\csname @twocolumnfalse\endcsname 
\title{Electron-Phonon or Hole Superconductivity in $MgB_2$?}
\author{J. E. Hirsch$^{a}$ and F. Marsiglio$^{b}$ }
\address{$^{a}$Department of Physics, University of California, San Diego,
La Jolla, CA 92093-0319\\
$^{b}$Department of Physics, University of Alberta, Edmonton,
Alberta, Canada T6G 2J1}

\date{February 27, 2001} 
\maketitle 
\begin{abstract} 
The BCS electron-phonon mechanism and the unconventional 'hole mechanism' have been
proposed as explanations for the high temperature superconductivity observed in
$MgB_2$. It is proposed that a critical test of which theory is correct is the 
dependence of $T_c$ on hole doping:  the hole mechanism predicts that $T_c$ 
will  drop rapidly to zero as holes are added, while the electron-phonon mechanism appears 
to predict increasing $T_c$ for a substantial range of hole doping.
Furthermore, the hole mechanism and electron-phonon mechanism differ qualitatively 
in their predictions of the effect on $T_c$ of change in the $B-B$ distances. We discuss 
predictions of the hole mechanism for a variety of observables
as a function of doping, emphasizing the expected differences and similarities
with the electron-phonon explanation. The hole mechanism predicts coherence length
and penetration depth to increase and decrease monotonically with hole doping respectively. 
\end{abstract}

\vskip2pc]
 
\section{Introduction}
Superconductivity at $40K$ in $MgB_2$ was not predicted by theory. After 
its discovery\cite{super}, it has been proposed that this finding is
expected within two fundamentally different theoretical frameworks:
the BCS electron-phonon theory\cite{elph,elph2} and the theory of hole
superconductivity\cite{hole}. Both theories have claimed to be consistent
with various experimental observations. The purpose of this paper is to expand on the
predictions of the theory of hole superconductivity, and to make sharper
the distinction between it and the electron-phonon theory $before$ critical
experiments are performed that can differentiate between both theories.
It is generally easier to differentiate between theories by comparing
their $predictions$, rather than their $postdictions$, of experimental
observations; elaborate theoretical frameworks can often find 
consistent explanations even for the most unexpected observations.

The electronic structure of $MgB_2$ is well established, as a variety of
old\cite{band1} as well as new\cite{elph,elph2,band2,band3} calculations are 
in essential agreement. Approximately $30\%$ of the density of states at
the Fermi energy is due to planar boron $p_{x,y}$ states ($\sigma$ bonds) 
that have little dispersion in the $z$ direction, giving rise to nearly
cylindrical hole Fermi surfaces of 2D character. The remaining $70\%$ of the density of
states originates in boron $p_z$ states ($\pi$ bonds) that are strongly
hybridized with the $Mg$ $s-p$ orbitals, have 3D character and give rise to mostly
electron-like Fermi surfaces. No $d$-electrons exist in either $Mg$ or $B$, so that 
magnetic and strong correlation mechanisms (generically called 'big tent' 
mechanisms\cite{bigtent})
proposed for the high $T_c$ cuprates, do not appear to be applicable. 

Within the electron-phonon framework, two different explanations have been
proposed, hereafter referred to as EP1\cite{elph} and EP2\cite{elph2}.
Both explanations emphasize the importance of strong bonding of the boron
atoms in giving rise to strong electron-phonon coupling, as well as the
light ionic mass giving rise to a large prefactor in the BCS-Eliashberg
expression for the transition temperature. They appear to differ in
the relative contribution of the boron states. Whereas EP1 appears to suggest
that contributions from all  states are important, EP2 attributes superconductivity
exclusively to the nearly full boron $p_{x,y}$ states. It is argued that the
observation of a boron isotope effect\cite{isotope} (isotope coefficient
$\alpha=0.29$) strongly favors electron-phonon mechanisms\cite{elph,elph2,isotope}.

In contrast, within the theory of hole  superconductivity\cite{hole1,hole2} the 
electron-phonon interaction is irrelevant, and instead superconductivity originates in 
undressing of hole carriers, driven by Coulomb interactions, in bands that are 
almost full. The superconducting condensation energy is kinetic, since paired 
carriers have lower effective mass than unpaired ones, and electron-hole
symmetry breaking is central to the physics. In $MgB_2$, the fact that large
parts of the Fermi surface are strongly hole-like, together with the fact that
the boron planes where the holes propagate are highly negatively charged,
are proposed to be the essential factors giving rise to high $T_c$\cite{hole}. The
existence of an isotope effect is generically expected within this
theory also\cite{hole,hole2}, although its magnitude is difficult
to calculate; a simple estimate yields\cite{hole} a much larger isotope effect than
observed experimentally in $MgB_2$ if the electron-phonon coupling
suggested in \cite{elph,elph2} is used.

Both the electron-phonon theory and the hole theory predict that the
superconducting state is s-wave, which appears to be supported by 
tunneling\cite{tun1,tun2,tun3,tun4} as well as NMR\cite{nmr}  measurements, 
and both theories are consistent with 
the observation of an isotope effect\cite{isotope}. The theory of hole superconductivity 
$requires$\cite{hole} that the conductivity in the normal state is hole-like, 
which is consistent with recently reported Hall effect measurements\cite{hall}; it
also $requires$\cite{hole}  superconductivity to disappear
 when the hole bands
in $MgB_2$ become full, which is consistent with reported experimental
results on $Mg_{1-x}Al_xB_2$\cite{alum}. Both of these facts apparently
are consistent with electron-phonon theory, as discussed in 
 EP2\cite{elph2}. On the other hand, the hole theory
predicts that $T_c$ should increase under pressure if the dominant effect
is reduction of the $B-B$ distances\cite{hole}, while electron-phonon theory 
(EP1\cite{elph}) predicts that pressure generically should reduce $T_c$;
experiments show that hydrostatic pressure reduces $T_c$\cite{chu,maple}, in
apparent agreement with electron-phonon theory. We return to
this point later in the paper.

A key prediction of both theories is the behavior of $T_c$ upon hole doping,
for example in the compound (not yet fabricated to our knowledge) $Li_xMg_{1-x}B_2$.
Here there is a clear opportunity for distinction between both theoretical frameworks.
EP1 explicitly states that decreasing the Fermi level 'may provide an
additional contribution to $\lambda$', which suggests that $T_c$ should increase
upon hole doping.
EP2 does not explicitly address this crucial point, but emphasizes the
'substantial value of the Fermi level density of states', even though
'the hole density $n_h$ is small'. Since the density of states increases
with hole doping, both for the  $\sigma$ as well as for the $\pi$ bands,
 for a very substantial range of hole doping, a reasonable inference 
 within electron-phonon theory
is that $T_c$ should also increase with hole doping for a substantial range.
Under the assumption that electron-phonon
matrix elements, phonon frequencies and Coulomb pseudopotential don't change substantially, 
electron-phonon theory  would 
predict that $T_c$ should increase with hole doping in a range
of about $2eV$ below the Fermi level of $MgB_2$ (following the density of states
increase), corresponding to $\sim 0.36$ holes
added per $B$ atom, and stay high (above the $T_c$ of $MgB_2$) till
about $4eV$ below $\epsilon_F$, corresponding to $\sim 0.61$
 holes added per $B$ atom. Instead, the theory of hole superconductivity
predicts that $T_c$ will rapidly drop when holes are added,
becoming small or zero before the number of added holes per $B$ atom
reaches only $0.12$. Comparison between these 
predictions of both theories will be discussed in detail in the next sections. 
No measurements on the behavior of $T_c$ with hole doping have yet been reported
to our knowledge.

\section{Model of hole superconductivity}

Within the model of hole superconductivity, as well as within EP2, the bands
that drive superconductivity are the nearly two-dimensional boron $p\sigma$
bands. Calculations for three-dimensional anisotropic band structures\cite{hole1}
have shown that a two-dimensional model reproduces the essential features,
hence we will ignore the third dimension here for the calculations with the model
of hole superconductivity. We will also approximate the nearly constant
density of states of the hole $p\sigma$ bands by a constant, which has
a negligible effect. The model is then defined by 4 parameters:
$U, K, W $ and $D$. $D$ is the bandwidth, $U$ the on-site Coulomb repulsion,
$W$ is proportional to the nearest neighbor Coulomb repulsion, and $K$ is
proportional to the correlated hopping interaction $\Delta t$ that drives 
superconductivity. The BCS pairing interaction is given by 
$V_{kk'}\equiv V(\epsilon_k,\epsilon_{k'})$, with
\beq
V(\epsilon, \epsilon')=U+2\frac{K}{D}(\epsilon+\epsilon')+
4\frac{W}{D^2}\epsilon\epsilon'.
\eeq
The critical temperature is determined by the equation
\beq
1=2KI_1-WI_2-UI_0+(K^2-WU)(I_0I_2-I_1^2)
\eeq
and the parameters $\Delta_m$ and $c$ that define the energy-dependent gap 
\beq
\Delta(\epsilon)=\Delta_m(-\frac{\epsilon}{D/2}+c)
\eeq
by the equations
\bmath
\beq
1=K(I_1+cI_0)-W(I_2+cI_1)
\eeq
\beq
c=K(I_2+cI_1)-U(I_1+cI_0)
\eeq
\emath
with
\bmath
\beq
I_l=\frac{1}{D}\int_{-D/2}^{D/2} d\epsilon (-\frac{\epsilon}{D/2})^l
\frac{1-2f(E(\epsilon))}{2E(\epsilon)}
\eeq
\beq
E(\epsilon)=\sqrt{(\epsilon-\mu)^2+\Delta(\epsilon)^2}
\eeq
\emath
with $f$ the Fermi function and $\mu$ the chemical potential.
Finally, the hole density $n_h$ is determined by the equation,
\beq
n_h = 1 - {2 \over D} \int_{-D/2}^{D/2} d\epsilon (\epsilon - \mu)
\frac{1-2f(E(\epsilon))}{2E(\epsilon)}.
\eeq

\section{Choice of parameters}
The density of states at the Fermi level of $MgB_2$ is estimated to be
0.75 states/eV, of which approximately 0.25 states/eV is ascribed
to the B $p\sigma$ states\cite{elph,elph2}. In a model with
constant density of states and a single band, that would correspond
to $D=4 eV$. There are two $p\sigma$ bands that contribute to 
this density of states, approximately with $2/3$ and $1/3$ weight\cite{elph2}
(heavy and light hole bands respectively in the nomenclature of
An and Pickett). According to An and Pickett, the equivalent two-dimensional
flat bands have densities of states 0.18 states/eV and 0.07 states/eV,
which would correspond to bandwidths $D_1=5.6 eV$, $D_2=14eV$ respectively. 
The total number of holes in the B $p\sigma$ bands for $MgB_2$ is
estimated to be $n_h=0.13$/unit cell, of which approximately $0.09$ and
$0.044$ holes are in the heavy and light hole bands respectively. The
total $n_h$ per B atom in $MgB_2$ is approximately $0.067$.

We will discuss elsewhere the results of our theory in the presence of two
hole bands, which we do not expect will be qualitatively different\cite{twoband}.
Here, we will use a single 'effective band' of bandwidth $D=5eV$, Coulomb
repulsion $U=5eV$ and nearest neighbor repulsion $W=0$. As we will discuss
in a later section, the model gives similar results for a wide range of
parameters. For these parameters, we choose the value of $K$ required to yield
$T_c=40K$ for $n_h=0.067$, which is $K=2.97 eV$. As discussed elsewhere\cite{hole},
we believe this value of $K$ is reasonable for $MgB_2$, but  emphasize
that here $K$  is a fitting parameter as we have not
obtained it from a first-principles calculation.

\section{$T_c$ versus doping}
As discussed in the introduction, no calculations of $T_c$ versus hole doping
with the electron-phonon model (EP1 or EP2) have yet been reported. It is
possible that such calculations may indicate large changes in the phonon frequencies,
electron-phonon matrix elements  or Coulomb pseudopotential with doping. 
In the absence of other information
however, we will assume that all these quantities are constant with doping,
 and calculate $T_c$ within
the electron-phonon model by the modified McMillan formula \cite{elph}
\beq
T_c=\frac{<\omega_{log}>}{1.2}e^{-1.02\frac{(1+\lambda)}
{\lambda-\mu^*-\mu^*\lambda}} 
\eeq
with $\mu^*=0.1$ and  $<\omega_{log}>=700K$\cite{elph}, and $\lambda$ proportional to the
density of states in the B $p\sigma$ band. Because the fractional contribution of the 
B $p\sigma$ bands to the total density of states
 is quite constant with doping (1/3),
this calculation should predict the results of both electron-phonon models
EP1 and EP2 under the assumption that electron-phonon matrix elements,
phonon frequencies and $\mu ^*$do not change with doping.

Figure 1 shows the results of this calculation for the two models, as function
of hole content in the $p\sigma$ bands per B atom, $n_h$. Note that 
$n_h$ is approximately 1/3 of the total hole doping per unit cell, $n_h^{tot}$. The
electron-phonon model predicts that $T_c$ will continue to increase well
beyond the point where the hole model predicts $T_c$ will have vanished;
the maximum $T_c$ of $94K$ occurs for $n_h=0.43$ per B atom, or
total hole doping of $MgB_2$ of approximately $n_h^{tot}\sim 1.1$ holes
per unit cell, corresponding to
bringing the Fermi level down approximately $2.1 eV$ from its position
in $MgB_2$. In contrast, the maximum $T_c$
in the hole model of $49K$ occurs for $n_h\sim 0.035$ per B atom,
corresponding to $electron$ doping of $MgB_2$ of approximately
$0.03$ electrons per B atom, or $0.09$ electrons per unit cell.

As mentioned above, these results could be modified in the electron-phonon model
if  there are substantial changes in some or all phonon frequencies, 
electron-phonon matrix elements, or Coulomb pseudopotential. In the hole model, 
some modification may
be expected if the contribution of the two B $p\sigma$ bands is taken into
account separately. In particular, assuming it is the heavy hole band that
dominates $T_c$, $T_c$ would go to zero in an even narrower range of
hole doping than indicated by Figure 1. Despite these caveats,
we believe the qualitative difference in the behavior predicted by the 
hole model and by the electron-phonon models depicted in Figure 1 is
robust.

\section{Pressure dependence of $T_c$}

In the hole model, a decrease in the $B-B$ intraplane distances should
strongly increase $T_c$. So far, only observations of changes in $T_c$ under
hydrostatic pressure on
polycrystalline samples have been reported\cite{chu,maple}, that indicate that such pressure
decreases $T_c$. We believe that hydrostatic pressure is likely to affect
much more strongly the lattice spacing in the $c$ direction than the planar
lattice spacings, due to the stiffness of the $p\sigma$ bonds.
Furthermore, it is possible that substantial charge transfer occurs between
different bands when pressure is applied. For example, in many high $T_c$ cuprates
the hole concentration in the planes is increased by approximately $10\%$ when
$1GPa$ hydrostatic pressure is applied\cite{schilling}. According to Figure 1, a $10\%$ increase
in hole content from $MgB_2$ leads to a decrease in $T_c$ of $3K$; the
reported observation of a decrease in $T_c$ by $1.6K$\cite{chu} could hence
be accounted for by such an increase in the B  $p\sigma$ orbitals hole content
together with a small decrease in the $B-B$ distances. Hence the observation is
not necessarily inconsistent with our model, as also emphasized in ref. \cite{chu}.

Here we consider the effect on $T_c$ of a decrease in the $B-B$ intra-plane
distances. The intrinsic effect of changing lattice spacing in the c direction
should be much smaller within our model\cite{pressure}. 
 In the electron-phonon model, we assume the dominant
effect will be to decrease the density of states. In the hole model, we
assume the effect is to increase the bandwidth (i.e. decrease the density of states)
and increase the interaction parameter $K$, that depends on overlap matrix
elements as the bandwidth does, by the same fraction.

Figure 2 shows the changes expected under a $5\%$ change in these parameters,
achieved by either physical or chemical pressure. In the electron-phonon
model (again assuming no change in phonon frequencies,
electron-phonon matrix elements and $\mu^*$), a small decrease in $T_c$ results .
 In the hole model, a strong increase
in the critical temperature for all hole dopings results. By performing such
experiments and monitoring the changes in lattice constants and in carrier
concentration (e.g. through Hall measurements) we hope it will be possible
to decide which of the two qualitatively different behaviors shown in
Figure 2 takes place in this class of materials.

\section{Other results for the model of hole superconductivity}

We next discuss other results for our model for a single band with
constant density of states, in the range of parameters that may be
appropriate for this class of materials. Figure 3 illustrates the
effect of changing Coulomb interaction parameters in the model, 
always choosing the parameter $K$ so as to yield the observed
value $T_c\sim 40$~K for $n_h=0.065$. It can be seen that the
behavior of $T_c$ versus doping is quite insensitive to large
variations in the Coulomb interactions. On increasing the nearest
neighbor repulsion the range of hole dopings where $T_c$ is nonzero 
increases somewhat, and on increasing the on-site Coulomb interaction
that range decreases. If the Coulomb repulsion appropriate for
$MgB_2$ is larger than $5$ eV, the maximum $T_c$ obtained by
electron doping could be larger than $50$ K, as seen in Figure 3.

Figure 4 illustrates the effect on $T_c$ of changing the bandwidth,
i.e. the density of states. It can be seen that the effect is again
remarkably small, with a reduction in the bandwidth leading to a small
decrease in the range of hole concentration where $T_c$ is non-zero.

We next calculate other observables for the parameters of Figure 4. 
Figure 5 shows the behavior of the coherence length
$\xi_o$, defined as the average size of the pair wave function, with
hole doping. The formulas to evaluate this quantity for the model
under consideration here are given in ref. \cite{strong}. The coherence
length is found to be almost independent of the values of the Coulomb
interactions, but it depends strongly on the bandwidth, as seen in
the different curves in Figure 5: as the bandwidth decreases, the
coherence length decreases. The coherence length in Figure 5 is given
in units of lattice spacings in an effective square lattice; to transform
to physical units for $MgB_2$ ($a=3.14 \AA$), a lattice spacing $a_{eff}=2.22$ \AA \, should be
used. For hole concentration $n_h=0.065$, the result obtained with
$D=2.5$ eV is close to the observed experimental value,
 $\xi_o=45$ \AA\cite{finne,budko}.
Instead, for $D=5$ eV we obtain a coherence length of $\xi_o=83$ \AA,
larger than seen experimentally. However our calculated value
corresponds to the in-plane coherence length which is not necessarily
the same as that measured in a polycrystalline sample. As seen in Figure
5, the coherence length is predicted to increase monotonically with hole doping.

Similarly, Figure 6 shows the behavior of the in-plane London penetration
depth $\lambda_L$, assuming the clean limit:
\beq
\lambda_l=4638(d(A))^{1/2}\frac{1}{T_a(meV)]^{1/2}}
\eeq
with $d$ the distance between boron planes, $d=3.52$ \AA, and 
$T_a$ the average in-plane kinetic energy per boron atom.
The estimated value for $MgB_2$  is $\lambda_L=1400$ \AA \cite{finne}, which is close to
the value given in Figure 6 for $D=2.5$ eV and $n_h=0.065$, $\lambda_L=1344$ \AA.
For $D=5$ eV we obtain a smaller value than seen experimentally. The penetration
depth is predicted to decrease monotonically with hole doping. In conjunction
with the increasing coherence length, this implies that the Ginzburg-Landau
parameter $\kappa=\lambda_L/\xi_o$ will rapidly decrease with hole
doping, which could eventually lead to a cross-over from type II to type I behavior
for high hole doping.
However this is likely to be prevented by disorder, that would cause an increase in the
penetration depth from its clean limit value.

The gap versus hole concentration follows closely the behavior of the
critical temperature. This is shown in Figure 7. In contrast to
high $T_c$ oxides, we do not find a substantial increase in the
gap ratio in the underdoped regime, because the parameters here correspond
to a weak coupling regime.
In Figure 8 we show the hole concentration
dependence of the specific heat jump at $T_c$, which agrees with the BCS
weak coupling value $1.43$ for high hole concentrations and becomes
larger for low hole concentrations, particularly as the bandwidth becomes
smaller.

The temperature dependence of various quantities obtained from our
model also follows closely
the BCS weak coupling behavior. As an example we show results for the
gap ratio and the specific heat for one parameter set in Figure 9.
Experimental results for specific heat of $MgB_2$ show a clear
specific heat jump at the transition with value close to the
expected BCS value\cite{spec}.

Finally Figure 10 shows tunneling characteristics for one set of parameters
and hole doping appropriate to $MgB_2$. Again the behavior resembles the
weak coupling BCS results for an s-wave gap, except for the existence of
asymmetry.
As emphasized elsewhere\cite{hole1}, an asymmmetry of universal sign
occurs for this model, with a larger peak for a 
negatively biased sample. The case of Figure 10 corresponds to the smallest
bandwidth considered; for larger bandwidth the magnitude of tunneling asymmetry
decreases.

\section{Conclusions}

The growing number of experimental results on $MgB_2$ suggests that superconductivity
in these materials is more akin to conventional superconductivity than
it is to high temperature superconductivity in the cuprates. Thus it is
natural that a consensus is growing that $MgB_2$ is describable within the
conventional BCS-electron-phonon framework. However, we have proposed in
ref. \cite{hole} that instead $MgB_2$ should be described by the model of 
hole superconductivity, just as the high $T_c$ cuprates\cite{hole1}. The
common elements in the two classes of materials are that conduction is dominated by 
carriers in nearly filled bands, i.e. of hole-like character, and that
the carriers that drive superconductivity propagate 
 in conducting substructures that are highly negatively charged (planes). 
Differences in the behavior of the two classes of materials arise within the model
of hole superconductivity from the fact that they are in different parameter regimes:
the high $T_c$ cuprates are in a substantially stronger coupling regime
(as indicated by the shorter coherence lengths), particularly for low 
hole doping\cite{note}. Compelling aspects of the theory of hole superconductivity are
that it can describe superconductivity in a wide range of coupling regimes,
and that it could be a universal theory of superconductivity for all materials\cite{hole2}.

In the weak coupling regime, the predictions of the model of hole superconductivity
are similar to those of conventional BCS theory, and hence to the
predictions of weak coupling electron-phonon BCS theory. Hence experimental
evidence for BCS behavior in e.g. temperature dependence of the
gap\cite{tun2}, or in tunneling characteristics\cite{tun1,tun2,tun3,tun4}
should not be taken to favor the electron-phonon model over the model
of hole superconductivity. The isotope effect, conventionally assumed to
favor the electron-phonon model, is also expected within the model of
hole superconductivity\cite{hole2}, and hence should also not be used to
differentiate between both models. An Eliashberg analysis of fine structure
in tunneling characteristics above the gap energy, that traditionally has been assumed to
be the strongest proof for the electron-phonon mechanism, has not yet
been performed for this material.

Here we have focused on two properties that show a clear difference in the 
electron-phonon and the hole model. One is the hole doping dependence of the critical
temperature, which the hole model predicts to be much stronger than the
electron-phonon model. These experiments have not yet
been performed, and once experimental
results become available it will be possible to ascertain which of both
models is favored. Of course it is possible that even if experiments show
that superconductivity is rapidly suppressed with hole doping, as the
hole model predicts, electron-phonon theory may also account for it if
a rapid decrease of electron-phonon matrix elements or of the relevant
phonon frequencies with hole doping is postulated to occur, or if
a rapid increase in the Coulomb pseudopotential $\mu^*$ 
with hole doping is postulated to occur. If so, electron-phonon
theory (e.g. in it's EP2 version) and the hole theory will become
increasingly indistinguishable.

The other property that shows a clear difference in the 
electron-phonon models and the hole model is
 the pressure dependence of $T_c$ for uniaxial pressure that
modifies the intra-plane $B-B$ distances: the hole model predicts a 
strong increase in $T_c$, and the electron-phonon model a (weaker) decrease
in $T_c$. Again,  once experimental
results become available it will be possible to decide which 
model is favored. However, here again it is possible that even if experiments show
that $T_c$ is strongly enhanced by reduction of the intra-plane
$B-B$ distance,  electron-phonon theory could account for it  if
a concomittant increase of electron-phonon matrix elements or of the relevant
phonon frequencies is postulated to occur or a concomittant decrease
of Coulomb pseudopotential is postulated to occur.

Assuming experimental results will show a rapid decrease of $T_c$ with
hole doping as predicted by our theory, it is interesting to examine
how the parameters in electron-phonon models would have to change to
account for such a drop, given the band structure results for the
density of states shown in Figure 1. Figure 11 shows three possible scenarios:
(1) the behavior required
for the average of the square of the electron-phonon matrix element
$<g^2>$ versus hole doping of the B atoms, assuming the phonon frequencies 
of the relevant phonons and the Coulomb pseudopotential $\mu^*$ stay
constant; (2) the behavior required for the phonon frequencies of the
relevant phonons $<\omega_{log}>$, assuming the electron-phonon matrix elements 
and $\mu^*$ stay constant, and (3) the behavior required
for the Coulomb pseudopotential $\mu^*$, assuming the
electron-phonon matrix elements and  phonon frequencies of the relevant phonons stay
constant. 
It can be seen that in all cases a rather rapid variation of parameters with
hole doping is required. Of course a suitable combination of decrease in electron-phonon
matrix elements and phonon frequencies and increase in $\mu ^*$
could also account for such behavior.
It should be stressed that such rapid variations of  electron-phonon matrix elements, 
 phonon frequencies and/or $\mu ^*$ with hole doping  have so far not been predicted
by the electron-phonon models\cite{elph,elph2}.

The range of hole doping where superconductivity occurs in our model is not
strongly dependent on the parameters in the model for a wide range of
parameters, as was shown in Figures 3 and 4. Hence the prediction that
superconductivity should only occur in a narrow range of doping around
$MgB_2$ is a strong prediction of the model. If the band structure results for the
position of the Fermi level are correct, it implies that $MgB_2$ is somewhat
overdoped in our model, and hence doping with electrons should increase $T_c$.
This is in apparent contradiction with experimental results for 
$Mg_{1-x}Al_xB_2$\cite{alum}. Possible explanations for the discrepancy may be
problems with sample quality, or that the lattice constants change with
increasing $Al$ content.

We have also examined here the predictions of the model of hole superconductivity
for various observables in a range of parameters that appears to be
appropriate for $MgB_2$. Because this is a weak coupling regime, most
properties are found to be very close to conventional BCS behavior. 
If the appropriate bandwidth is rather small the universal asymmetry in
tunneling predicted by the theory becomes of appreciable magnitude.
We found the penetration depth decreasing monotonically with hole doping (assuming the
clean limit), which is the same qualitative behavior seen in high $T_c$ 
cuprates\cite{uchida}. The coherence length was found to increase
monotonically with hole doping, which is also seen in high $T_c$ materials.
If experiments confirm these predictions they will support the proposed
commonality in the physics of superconductivity in the cuprates
and in $MgB_2$-derived compounds. 
Given this behavior, in the absence of disorder a crossover from type II to type I 
behavior with hole doping should eventually occur. While disorder is likely to
prevent this in alloys, we expect that if similar stochiometric compounds
with larger hole content than $MgB_2$ are found they will have a smaller 
Ginzburg Landau parameter $\kappa$ than $MgB_2$ ($\kappa\sim 26$)\cite{finne} 
and possibly even be type I. 

Superconducting properties as function of hole doping have not yet been
discussed within electron-phonon models. We note however that the
strong increase of density of states expected with hole doping suggests that
electron-phonon theory may describe a crossover to a stronger coupling
regime with hole doping, i.e. decreasing coherence length and
increasing penetration depth. We stress that this would be in 
qualitative disagreement with our predictions.

In future work we will examine the predictions of our model taking into account
the presence of two different hole bands, and the effect of the
anisotropic band structure, in order to calculate observables in
different directions which will be of interest once experimental results in single
crystals become available.

\acknowledgements
FM was supported by the Natural Sciences and Engineering
Research Council (NSERC) of Canada and the Canadian Institute for Advanced
Research.

\begin{figure}
\caption {Comparison of the predictions for variation of the critical temperature
with hole doping in the model of hole superconductivity (full line) and the
electron-phonon model (dashed line). Here and in the following
figures, $n_h$ is the average hole content per Boron atom; the total hole
doping per unit cell is approximately three times larger. The results for
electron-phonon theory were obtained assuming constant electron-phonon matrix
elements and phonon frequencies, and using the density of states values
obtained in the band structure calculation in Ref. 2.
}
\label{Fig. 1}
\end{figure}

\begin{figure}
\caption { Comparison of the predictions for variation of the critical temperature
with in-plane B-B distance  in the model of hole superconductivity (full lines) and the
electron-phonon model (dashed lines). The bandwidth and density of states are assumed to 
change by  $5\%$. Again, the results for
electron-phonon theory were obtained assuming constant electron-phonon matrix
elements and phonon frequencies under compression.
}
\label{Fig. 2}
\end{figure}

\begin{figure}
\caption { $T_c$ versus hole concentration for bandwidth $D=5eV$ and
various values of Coulomb interaction 
parameters, given in the Figure in $eV$. 
The values for the correlated hopping parameter used for the
four cases shown are in $eV$, $K=2.97, 4.115, 5.135, 4.134$   respectively, in the order 
given in the Figure label.
}
\label{Fig. 3}
\end{figure}

\begin{figure}
\caption { $T_c$ versus hole concentration for on-site Coulomb repulsion
$U=5eV$ $W=0$, and various values of the bandwidth $D$, given in the Figure in $eV$. 
The values for the correlated hopping parameter used for the
three cases shown are in $eV$, $K=2.97, 2.496,1.963$   respectively, in the order 
given in the Figure label.
}
\label{Fig. 4}
\end{figure}

\begin{figure}
\caption {Coherence length (in units of the lattice spacing) versus doping 
for the three sets of parameters of
Figure 4. A lattice spacing corresponds to 2.2 \AA.
}
\label{Fig. 5}
\end{figure}

\begin{figure}
\caption {London penetration depth versus doping for the three sets of parameters of
Figure 4. 
}
\label{Fig. 6}
\end{figure}

\begin{figure}
\caption { 
Gap versus hole doping for the three sets of parameters
given in Figure 4.
}
\label{Fig. 7}
\end{figure}

\begin{figure}
\caption {Specific heat jump versus doping for the three sets of parameters of
Figure 4. 
}
\label{Fig. 8}
\end{figure}

\begin{figure}
\caption { Temperature dependence of specific heat (a) and of energy gap (b)
for the parameters of Figure 1 and doping $n_h=0.065$.
}
\label{Fig. 9}
\end{figure}

\begin{figure}
\caption { Tunneling characteristics for the parameters of Figure 4
with $D=1eV$, doping $n_h=0.065$,
and temperatures $T/T_c=0.99, 0.9, 0.3, 0.1$.
Note the higher peak when the sample is negatively biased. For larger 
bandwidths the magnitude of the asymmetry decreases.
}
\label{Fig. 10}
\end{figure}

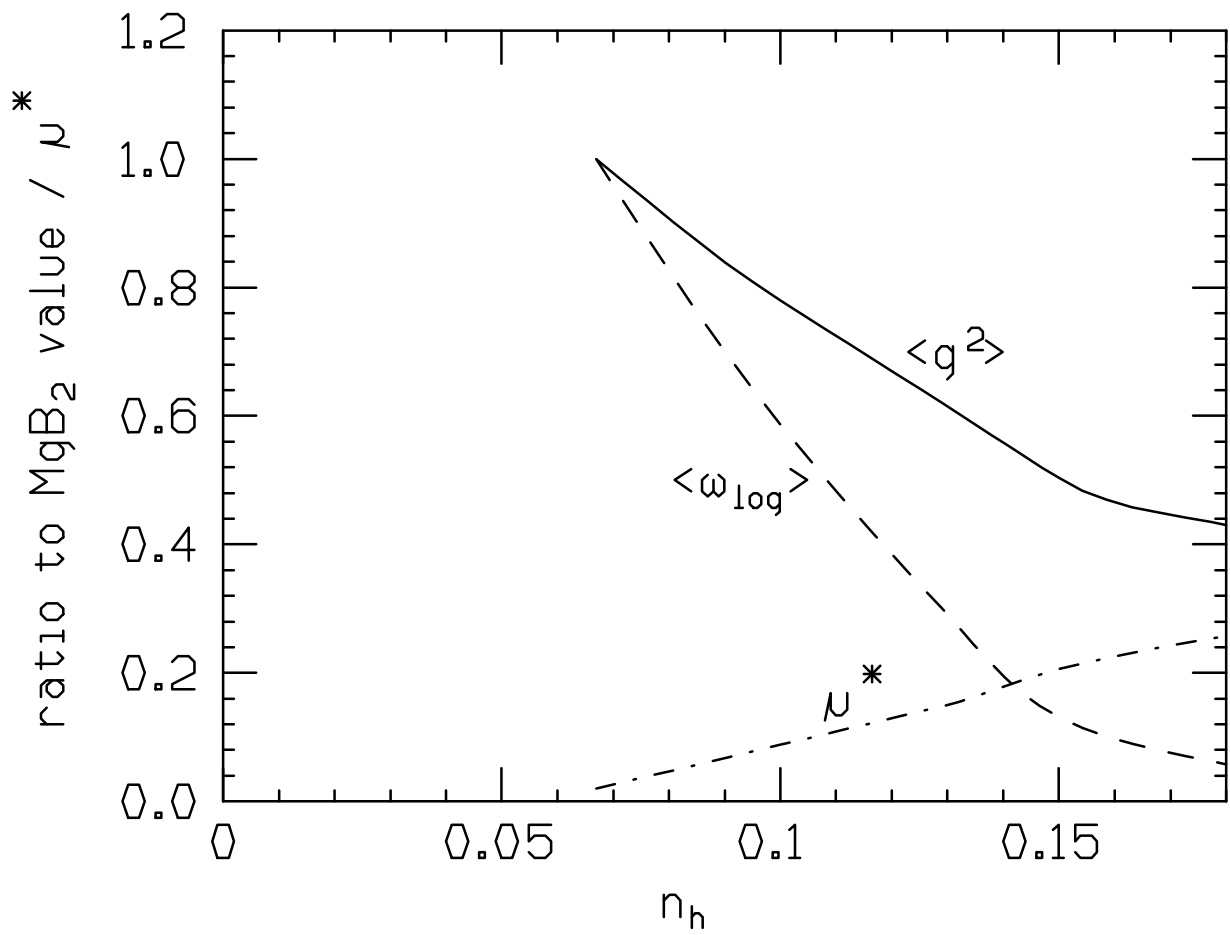
\begin{figure}
\caption { Variation of the electron-phonon matrix element $<g^2>$ (solid line), 
the average phonon frequency $<\omega_{log}>$ (dashed line),
and the Coulomb pseudopotential $\mu ^*$ (dash-dotted line) required, 
if $T_c$ is given by the electron-phonon model Eq. (7),
to yield the
rapid drop of $T_c$ with hole doping predicted by the model of hole
superconductivity. In each case it is assumed that the other parameters are 
fixed at their values for $MgB_2$. Density of states values obtained from
Ref. 2 are used.
}
\label{Fig. 11}
\end{figure}


\begin{references}
\bibitem{note} Furthermore in the cuprates additional phenomena and complications arise 
due to the presence of antiferromagnetism nearby in the phase diagram,
which we have argued to be irrelevant to
the physics of superconductivity of these materials\cite{hole1}.
\end{references}
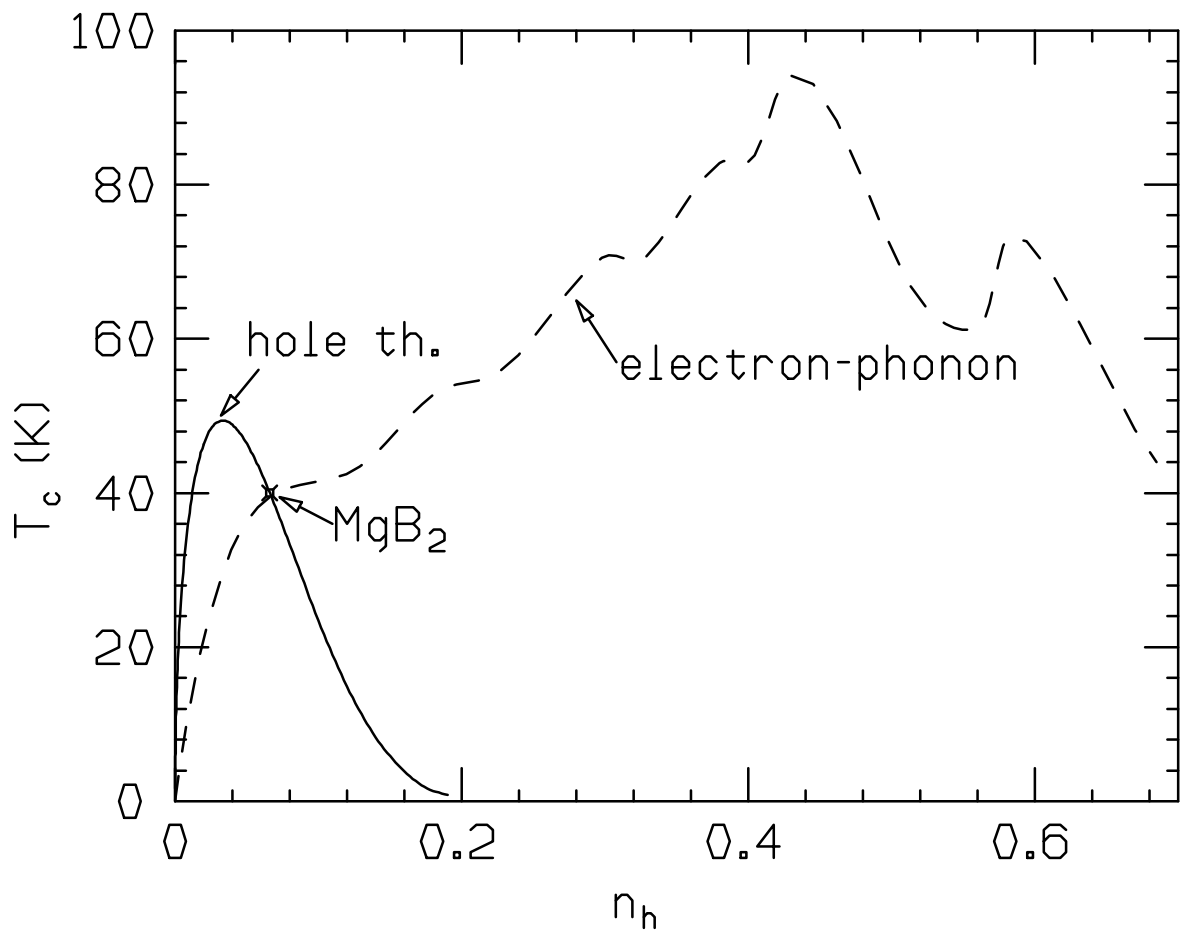
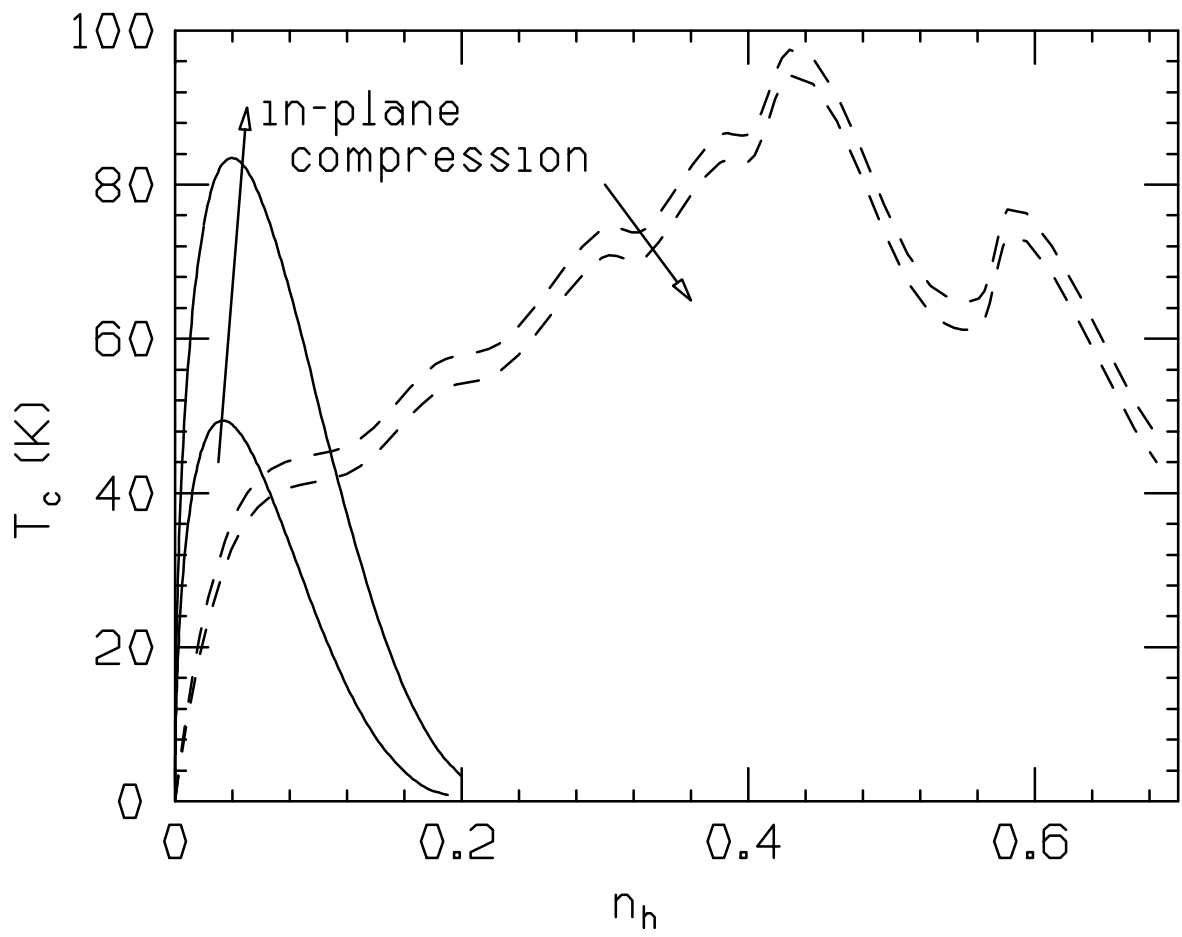
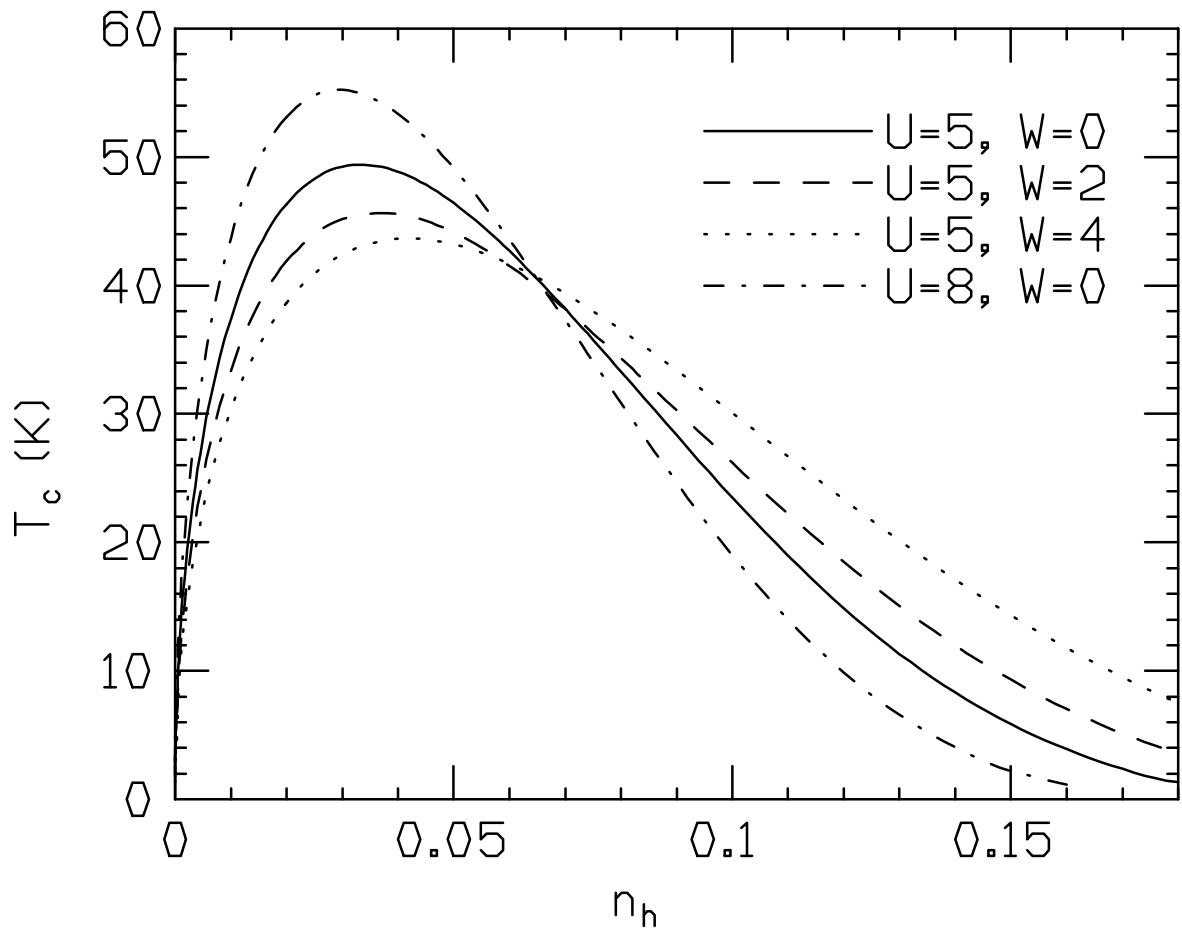
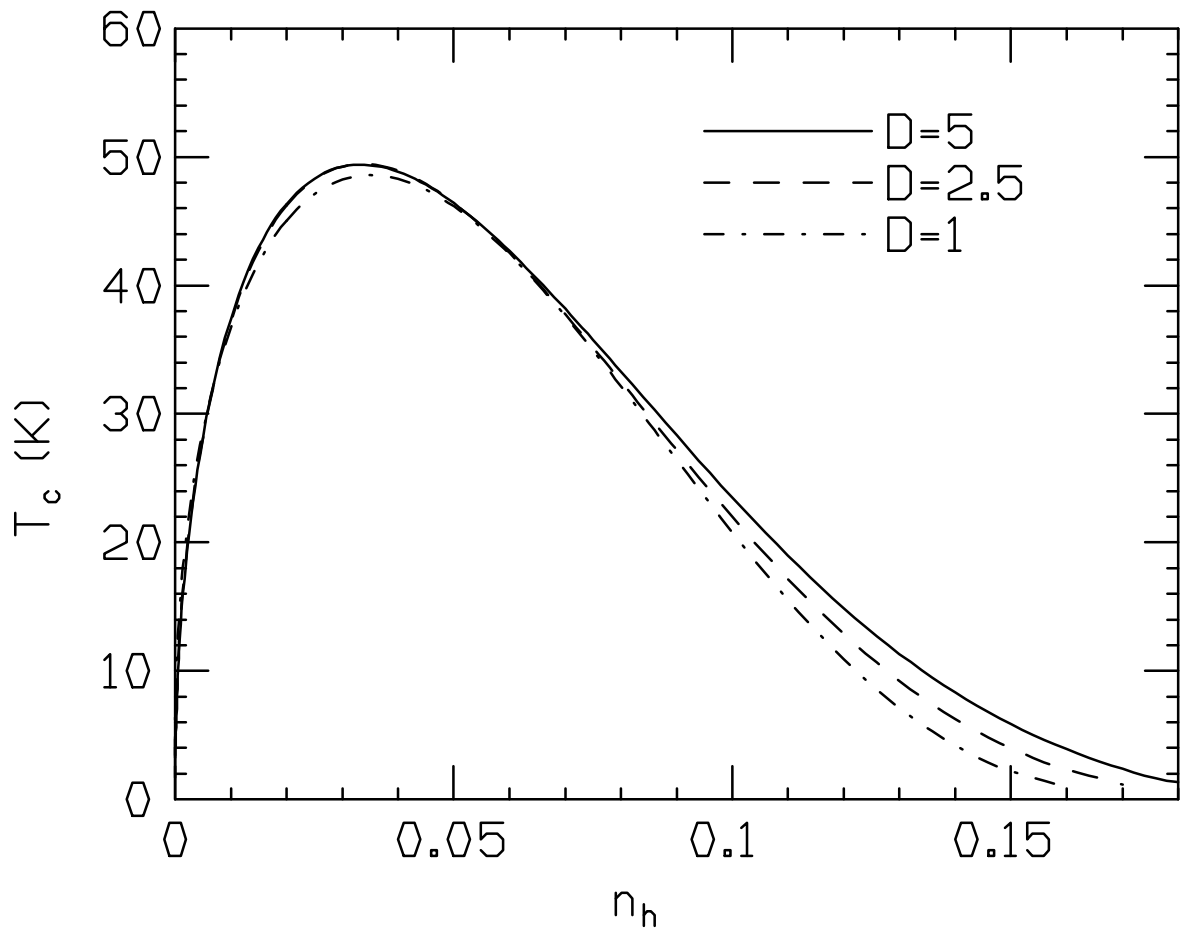
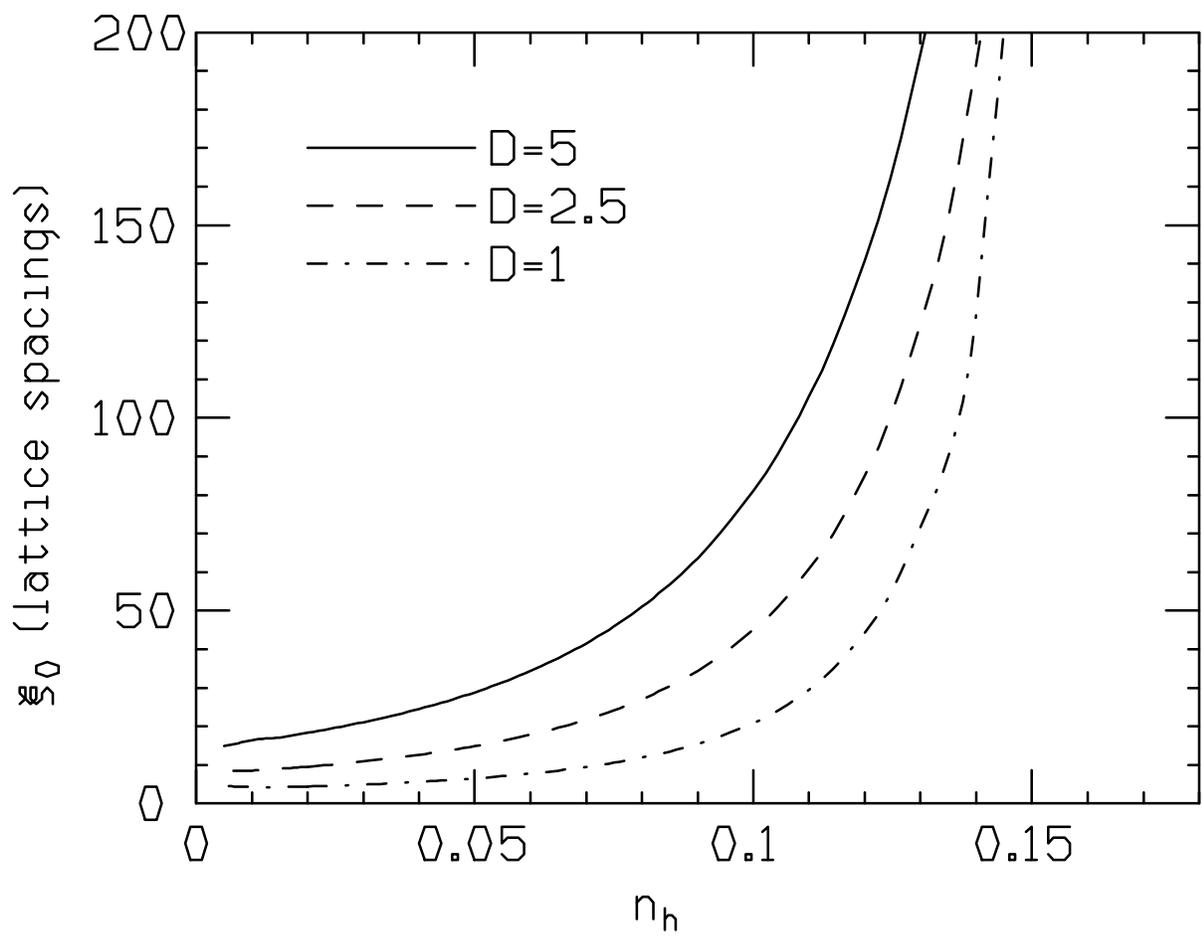
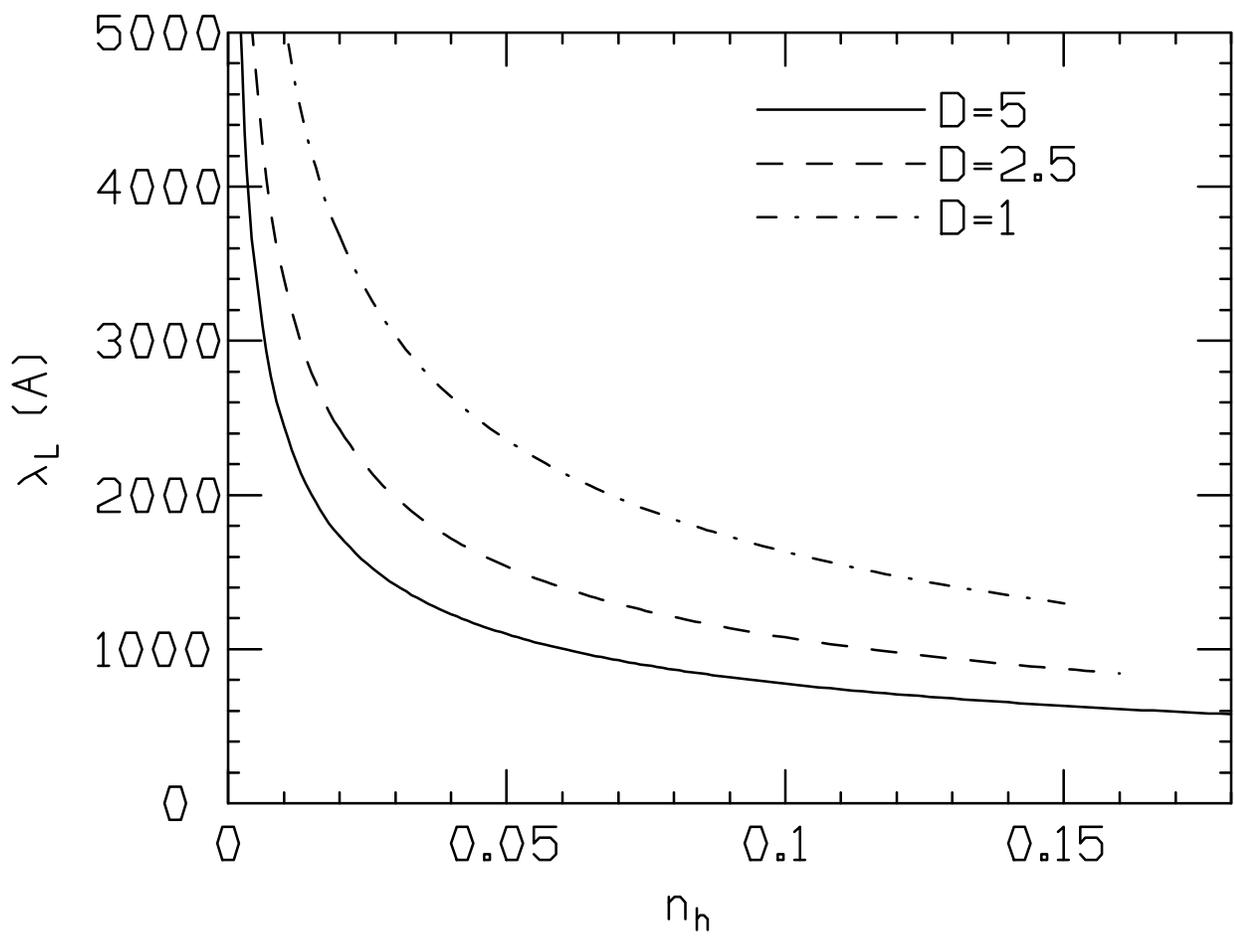
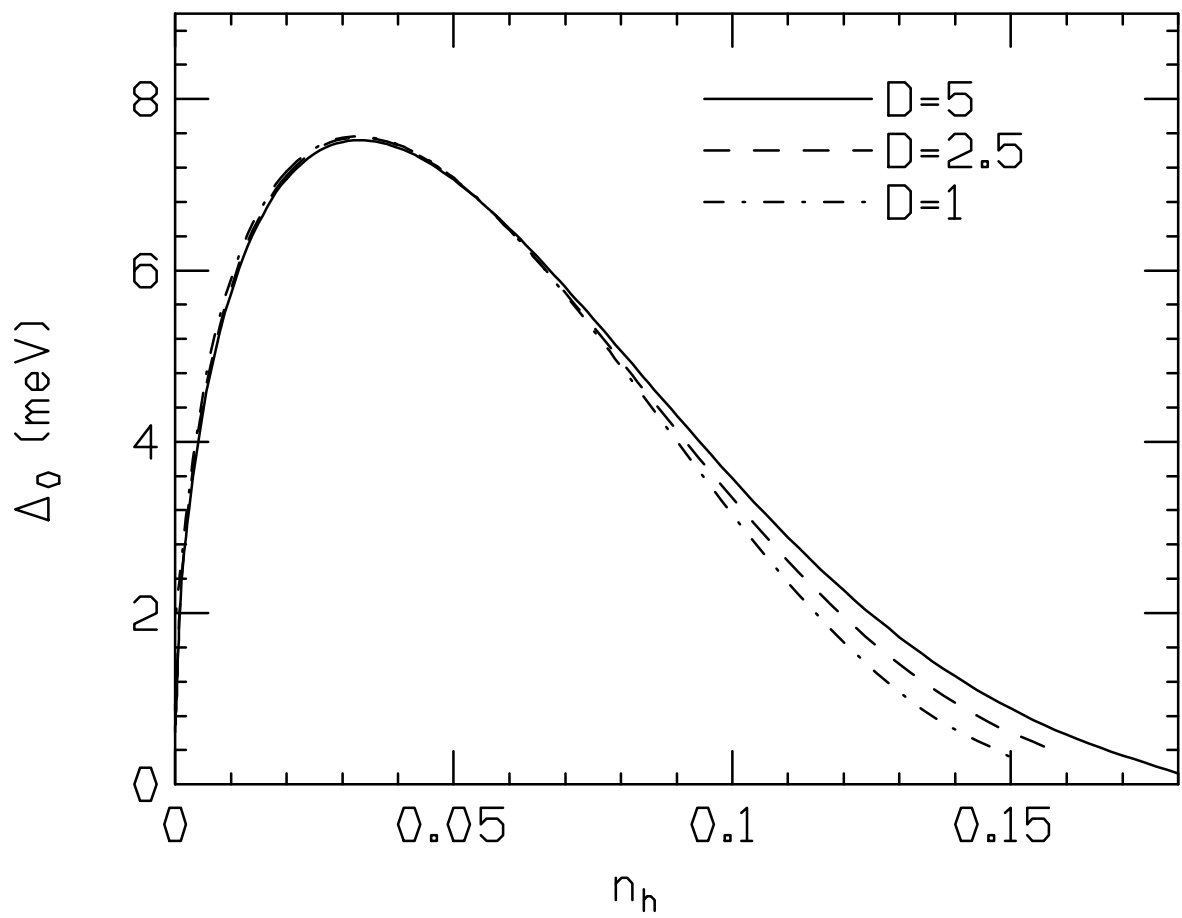
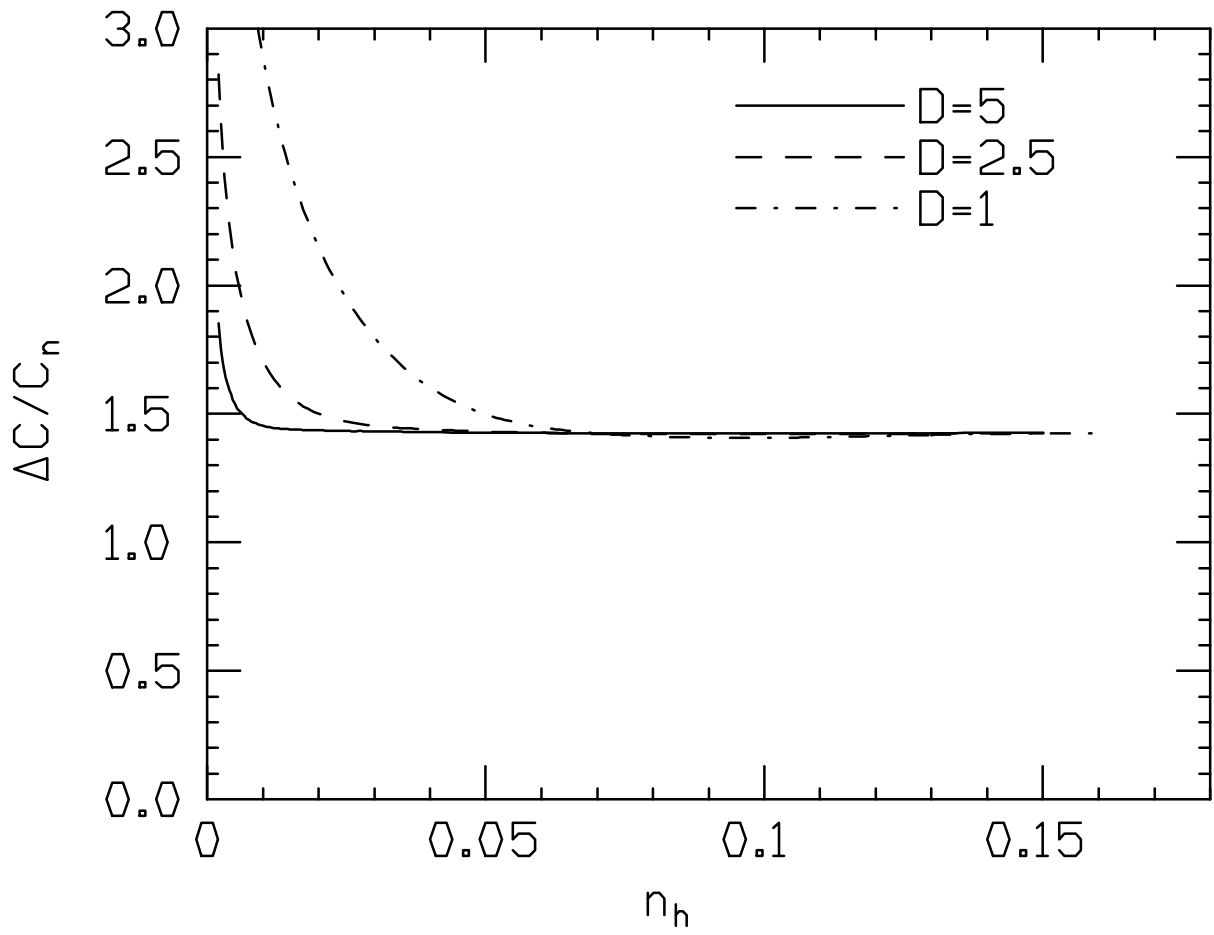
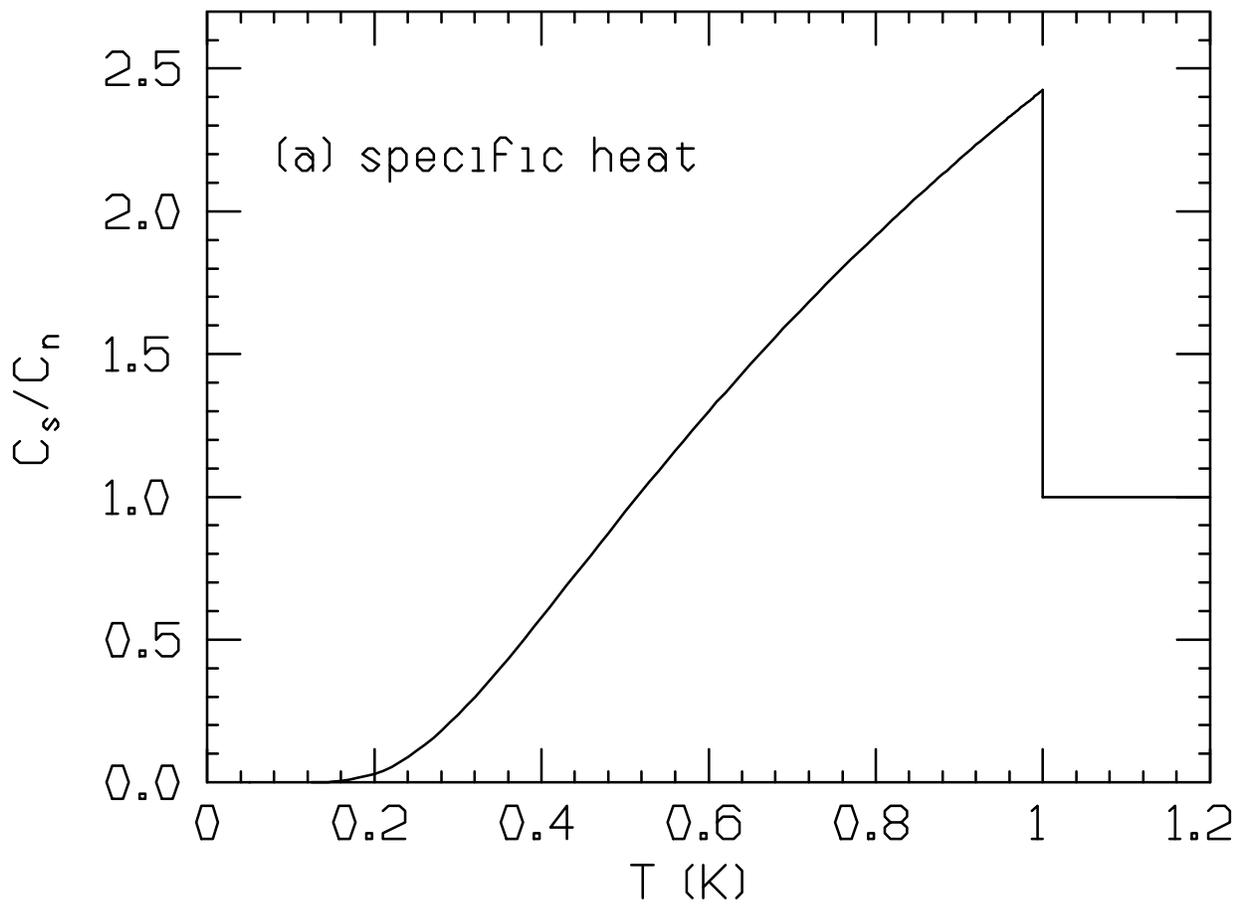
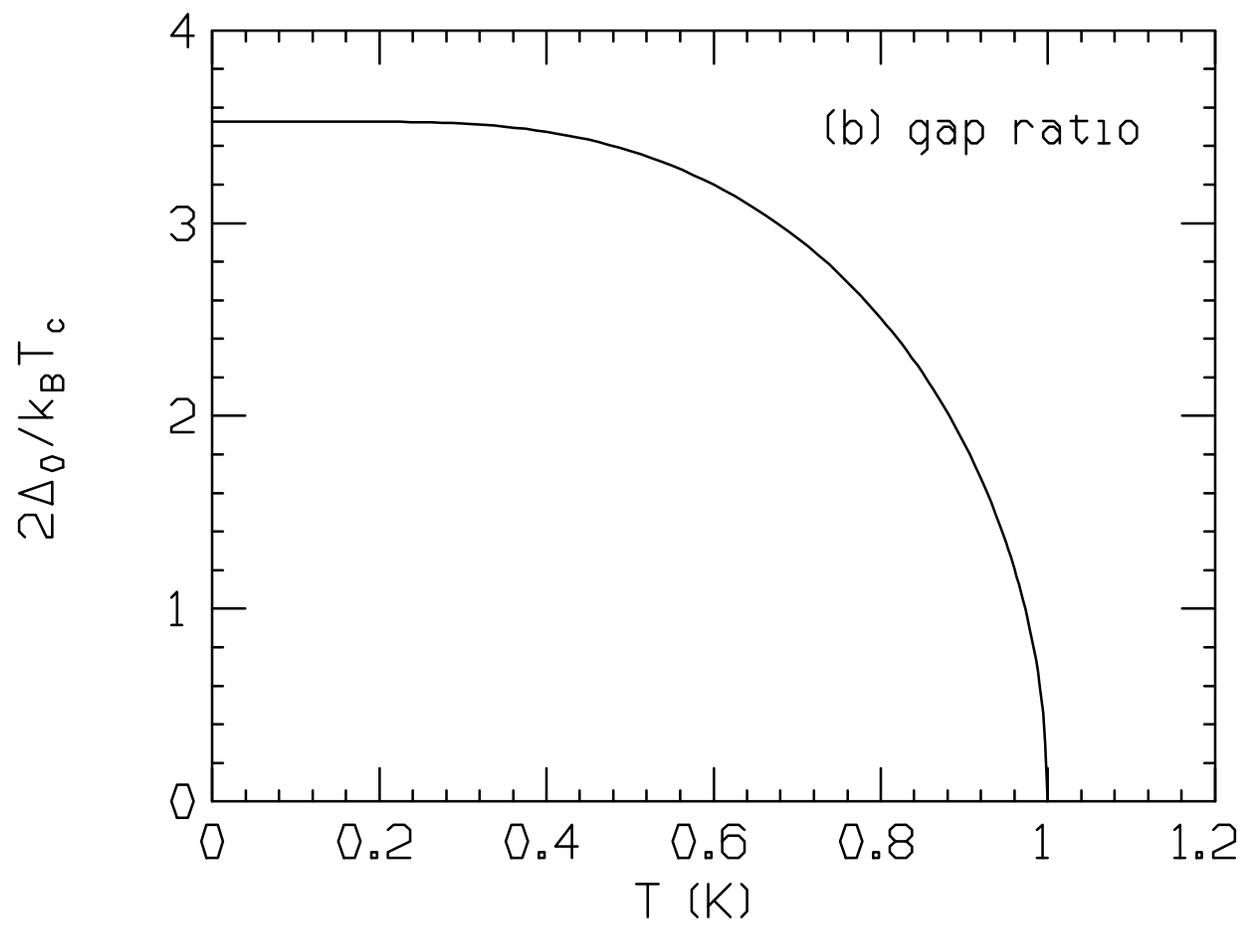
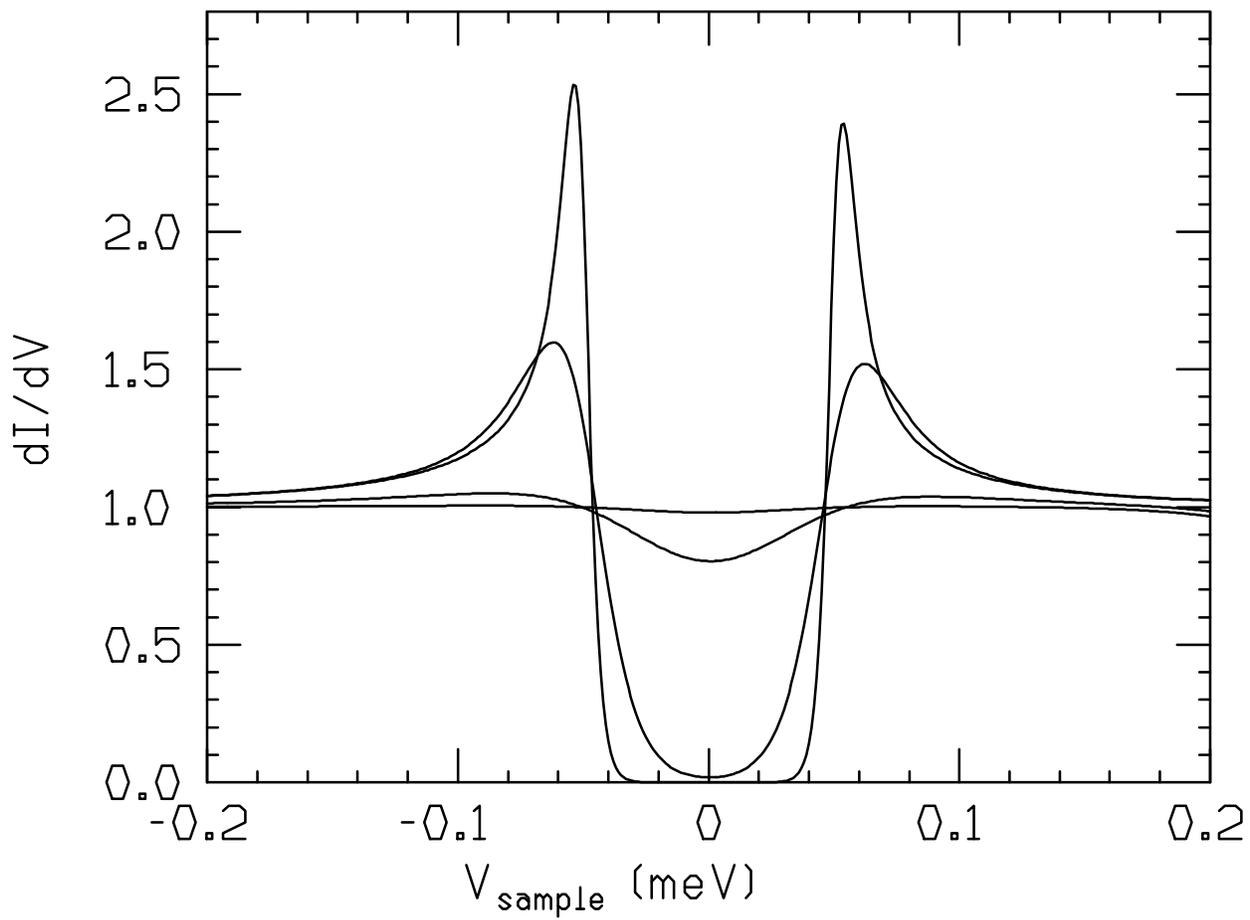
\end{document}